\documentclass[aps,prl,twocolumn,superscriptaddress,showpacs]{revtex4-1}

\usepackage{graphicx}

\newcommand{\OD}{\mathrm{OD}} 
\newcommand{\parmode}{{\bf e}_{\|}} 
\newcommand{\perpmode}{{\bf e}_{\bot}} 
\newcommand{\psiin}{{\bf E}_{\rm{in}}} 
\newcommand{\psiout}{{\bf E}_{\rm{out}}}

\begin{document}

\title{Dispersive Optical Interface Based on Nanofiber-Trapped Atoms}
\author{S. T. Dawkins}
\affiliation{Institut f\"ur Physik, Johannes Gutenberg-Universit\"at Mainz, 55099 Mainz, Germany}
 \author{R. Mitsch}
 \author{D. Reitz}
\affiliation{Institut f\"ur Physik, Johannes Gutenberg-Universit\"at Mainz, 55099 Mainz, Germany}
 \affiliation{Vienna Center for Quantum Science and Technology, TU Wien - Atominstitut, Stadionallee 2, 1020 Wien, Austria}
 \author{E. Vetsch}
 \affiliation{Institut f\"ur Physik, Johannes Gutenberg-Universit\"at Mainz, 55099 Mainz, Germany}
 \author{A. Rauschenbeutel}
\email{arno.rauschenbeutel@ati.ac.at}
\affiliation{Institut f\"ur Physik, Johannes Gutenberg-Universit\"at Mainz, 55099 Mainz, Germany}
\affiliation{Vienna Center for Quantum Science and Technology, TU Wien - Atominstitut, Stadionallee 2, 1020 Wien, Austria}

\date{\today}

\begin{abstract}

We dispersively interface an ensemble of one thousand atoms trapped in the evanescent field surrounding a tapered optical nanofiber. 
This method relies on the azimuthally-asymmetric coupling of the ensemble with the evanescent field of an off-resonant probe beam, transmitted through the nanofiber. 
The resulting birefringence and dispersion are significant; 
we observe a phase shift per atom of $\sim$\,1\,mrad at a detuning of six times the natural linewidth, corresponding to an effective resonant optical density per atom of $0.027$. 
Moreover, we utilize this strong dispersion to non-destructively determine the number of atoms. 

\end{abstract}
\maketitle

We have recently demonstrated a new technique for trapping and optically interfacing cold atoms~\cite{Vetsch2010}. 
Our method employs one-dimensional arrays of laser-cooled atoms trapped in a two-color evanescent field surrounding an optical nanofiber. 
The resulting atomic ensemble is both well-isolated from perturbations by the environment and efficiently coupled to a fiber-guided probe field. 
This makes our system a prime candidate for interfacing and manipulating trapped atoms with light. 

In~\cite{Vetsch2010}, the detection of cesium atoms was achieved by monitoring the transmission of resonant probe light through the nanofiber. 
This probe light couples efficiently to the atoms via its evanescent field resulting in an absorbance per atom of the order of one percent. 
This strong absorbance also implies that there is a significant phase shift of the probe light in the dispersive regime. 
In this paper, we present experimental evidence of this phase shift and show that it leads to a frequency-dependent birefringence that acts on the polarization state of the probe light propagating through the fiber. 

Being based on dispersive detection, our method has significant advantages over absorption or fluorescence-based techniques~\cite{Ketterle1999}. As an example, its signal-to-noise ratio is superior in the case of high optical depth when assuming shot-noise-limited detection~\cite{Lye2003}.
Conceptually, it is similar to other dispersive detection schemes for atoms and molecules such as interferometry~\cite{Petrov2007,Lodewyck2009}, frequency modulation spectroscopy~\cite{Savalli1999}, or phase-contrast imaging~\cite{Andrews1996}. 

In all these approaches, the phase shift induced by the atomic medium on the probe beam is compared to the phase of a reference beam via interference. 
In the case of atoms trapped using a nanofiber, this can be accomplished by interfering two orthogonal polarization modes, which couple unequally to the atomic ensemble. 
The polarization state of the output light thus enables one to infer the phase shift caused by the atoms. 

\begin{figure}
\includegraphics[width=0.45\textwidth]{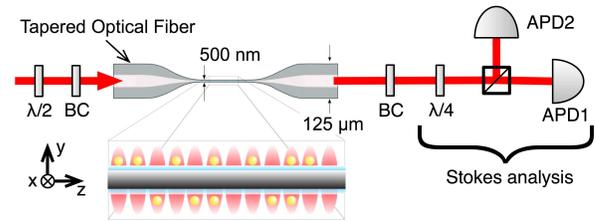}
\caption{ \label{fig:setup}
Schematic of the setup: An off-resonant laser beam is coupled into the nanofiber to probe the cesium atoms, which are trapped in the evanescent field of the nanofiber forming two one-dimensional arrays above and below the fiber (zoomed inset). A Stokes measurement is performed on the outgoing probe beam using a quarter-waveplate, a polarizing beam splitter, and two avalanche photodetectors (APD). A Berek compensator (BC) is used before and after the fiber to eliminate parasitic birefringence. 
  }
\end{figure}

Figure~\ref{fig:setup} shows a schematic of the experimental setup. 
The atoms are trapped in two one-dimensional arrays above and below the nanofiber waist of a tapered optical fiber~\cite{Vetsch2010}.  
For simplicity, the trapping fields are only schematically depicted in the zoomed inset. 
A probe beam is coupled into the tapered optical fiber with the linear polarization axis adjusted to $45^{\circ}$ with respect to the plane containing the atoms. 
Its initial polarization state in the nanofiber waist, $\psiin$, thus corresponds to an equal superposition of the two eigenmodes $\parmode$ and $\perpmode$, where $\parmode$ is the normalized fundamental HE$_{11}$ mode~\cite{Kien2004}, 
quasi-linearly polarized in the plane of the atoms ($y$-$z$-plane in fig.~\ref{fig:setup}), and $\perpmode$ is orthogonal to that. 
In this eigenmode basis,  the input light is thus described by 
$\psiin=\frac{\mathcal{E}_0}{\sqrt{2}} (\parmode + \perpmode) $, where $\mathcal{E}_0$ is the field amplitude. 
The light then propagates through the nanofiber, interacts with the atomic ensemble, and exits the other end of the fiber in an altered polarization state. 
The $S_3$ component of the Stokes vector describing this polarization state is then determined using polarization optics. 

\begin{figure}
\includegraphics[width=0.45\textwidth]{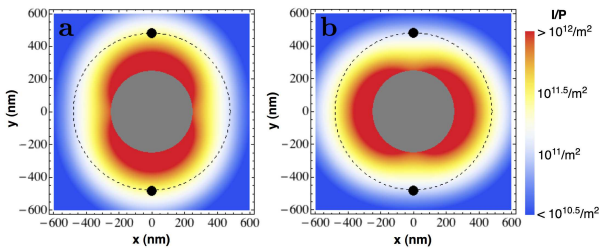}
\caption{ \label{fig:Polarizationmodes}
Evanescent intenstity distribution of the quasi-linearly polarized HE$_{11}$ modes, (a) $\parmode$  and (b) $\perpmode$. 
The filled black circles indicate the position of the trapped atoms at a distance of 230\,nm (black dashed circle) from the surface of the 500-nm diameter nanofiber (represented by gray cross-section). 
The intensity $I$ is scaled by total power $P$ and the color scale is logarithmic. 
}
\end{figure}

The origin of the difference between $\parmode$ and $\perpmode$ in coupling to the atoms derives from a twofold breaking of the azimuthal symmetry (see fig.~\ref{fig:Polarizationmodes}(a) and fig.~\ref{fig:Polarizationmodes}(b), respectively).  
The azimuthal symmetry of the propagation medium is broken by the presence of the arrays of trapped atoms above and below the nanofiber. 
At the same time, the intensity distribution of the modes exhibits an azimuthal dependence which is fixed by their polarization axis~\cite{Kien2004}.   
At a wavelength of 852\,nm, $\parmode$ couples 2.8 times more strongly to the atoms than $\perpmode$ 
because of the different respective intensities at the position of the trapped atoms (see fig.~\ref{fig:Polarizationmodes}). 
As a consequence, these eigenmodes become non-degenerate with regard to their propagation constants, thus giving rise to a birefringent effect. 
We note that our method is similar to conventional polarization spectroscopy, with the exception that the birefringence stems from the anisotropy of the atom-optical system, rather than being induced via optical pumping and the subsequent polarization of the atomic sample~\cite{Smith2004}. 

The resulting field amplitude of the light after interaction with the atoms is described by 
\begin{equation}
\psiout  =  
\frac{\mathcal{E}_0}{\sqrt{2}}  ( t_{\|} e^{i \phi_{\|}} \parmode+  t_{\bot} e^{i \phi_{\bot}}\perpmode),
\label{Eq_jonesforatoms}
\end{equation}
where the light--atom interaction is modeled by amplitude transmissions $t_{\|}$ 
and $t_{\bot}$ 
and phase shifts $\phi_{\|}$ and $\phi_{\bot}$, experienced by the eigenmodes $\parmode$ and $\perpmode$, respectively.  
Here, we assume that the birefringence of the fiber itself and other optical components is negligible (see below). 

Experimentally, the polarization state of the output light is analyzed with a polarizing beam splitter (PBS) in conjunction with a set of waveplates. 
By measuring with three different configurations of the waveplates, one can fully characterize the beam's polarization state and obtain the Stokes vector $S=\{ S_0, S_1, S_2, S_3 \} $~\cite{Saleh1991}. 
Here, we measure $S_3$ normalized to the total beam intensity, i.e., $S_3/S_0$, by inserting a quarter-waveplate with its axis aligned at 45$^{\circ}$ to the analyzing axis of the PBS.  
In terms of the powers at the output ports of the PBS, $P_{\sigma^+}$ and $P_{\sigma^-}$, we have 
\begin{equation}
\frac{S_3}{S_0} =  \frac{P_{\sigma^+}-P_{\sigma^-}}{P_{\sigma^+}+P_{\sigma^-}} = \frac{2 t_{\|}t_{\bot}}{t_{\|}^2+t_{\bot}^2} \sin{(\phi_{\|}-\phi_{\bot})}.
\label{Eq_stokes}
\end{equation}
The second equality in eq.~(\ref{Eq_stokes}) results from $P_{\sigma^+}=| {\bf e}_{\sigma^+}^* \cdot \psiout |^2$ and $P_{\sigma^-}=|  {\bf e}_{\sigma^-}^* \cdot \psiout |^2$, where ${\bf e}_{\sigma^{\pm}} =\frac{1}{\sqrt{2}}(\parmode \pm i \perpmode )$. 
In the following, we use $S_3/S_0\simeq \sin{(\phi_{\|}-\phi_{\bot})}$, where we have set the prefactor in the right-hand side of eq.~(\ref{Eq_stokes}) to unity. 
This is justified for $t_{\|}^2 \gtrsim 0.75$, i.e., less than 25\,\% absorption, where the errors of the approximation remain smaller than 1\,\%.

In order to deduce the absolute phase shift of $\parmode$, we assume that the phase shift is proportional to the atom--light coupling strength, which in turn is proportional to the intensity of the light at the position of the atoms.  
This assumption is valid as long as the ground state population is evenly distributed over all Zeeman sub-levels of the $F=4$ manifold, i.e., the ensemble has not been optically pumped. 
This is typically justified for atoms loaded from a magneto-optical trap~\cite{Metcalf1999}. 
The ratio is then $\phi_{\|}/\phi_{\bot}\simeq\,2.8$~\cite{Kien2004}, leading to 
\begin{equation}
\phi_{\|} = (1-\phi_{\bot}/\phi_{\|})^{-1} \Delta\phi \simeq 1.6 \Delta\phi=1.6\sin^{-1}(S_3/S_0),
\label{Eq_measurephaseshift}
\end{equation}
where $\Delta\phi=\phi_{\|}-\phi_{\bot}$ is the phase difference. 

The approach outlined above has been implemented using the apparatus described in~\cite{Vetsch2010} in conjunction with the measurement scheme laid out in fig.~\ref{fig:setup}. 
Berek compensators were used to eliminate any parasitic birefringence along the optical path, the majority of which is believed to arise in either the bulk fiber or the fiber taper. 
By observing the extinction ratio of the Rayleigh scattering from the nanofiber, we estimated the polarization impurity of the HE$_{11}$ mode of all employed fields to be at most 10\,\% percent. 
Finally,  $S_3/S_0$ is determined in accordance with eq.~(\ref{Eq_stokes}) using  two avalanche photodiodes (APDs) at the PBS output ports.

The correct alignment of the quarter-waveplate with respect to the PBS is achieved by distributing the power evenly to the APDs to give $S_3/S_0=0$ in the case where no atoms are trapped. 
Introducing trapped atoms then alters the polarization and changes the value of the Stokes measurement. 
In order to orient the input polarization at 45$^\circ$ with respect to $\parmode$, we adjust the angle of the input polarization until the measurement of $S_3/S_0$ in the presence of atoms yields approximately zero and thereby identify the orientation of $\parmode$ and $\perpmode$ for reference. 

\begin{figure}
\includegraphics[width=0.45\textwidth]{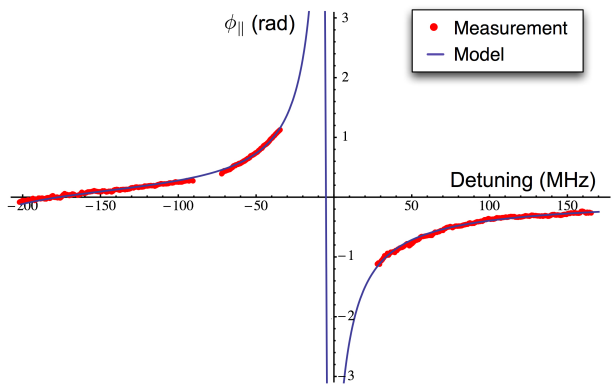}
\caption{ \label{fig:phasemeasurement}
Phase shift  $\phi_{\|}$ of the eigenmode $\parmode$ induced by 1000 atoms and measured as a function of the detuning from the $F=4 \rightarrow F'=5$ free-space transition frequency, $(\omega-\omega_{5})/2 \pi$, with 128 averages (red dots). The blue line is a fit using eq.~(\ref{Eq_phi}) taking into account the two nearest transitions $F=4 \rightarrow F'=4$ and $F=4 \rightarrow F'=5$. 
The former gives rise to the zero-crossing of the signal near $-$180\,MHz.}
\end{figure}

Figure~\ref{fig:phasemeasurement} shows the phase shift $\phi_{\|}$, measured as a function of the detuning of the probe with respect to the $F=4\rightarrow F'=5$ free-space transition frequency in cesium.  
As expected, the measurement yields a dispersive resonance signal which closely matches the theoretical model. 
We observe a good signal-to-noise ratio and phase shifts of more than 1\,rad at a detuning of 30\,MHz.

In this measurement, the probe laser was locked to various transitions of the cesium D2 line via polarization spectroscopy on a vapor cell and an acoustic optic modulator (AOM) was used to shift and scan the probe beam appropriately (scan duration was 0.5\,ms). 
The missing sections correspond to regions of poor signal due to being either in the absorptive regime or beyond the efficient range of the AOM. 
The mean atom number for this measurement was estimated at $N_{\rm{at}} =(1021\pm64)$ from the power absorbed by the ensemble at saturation in a series of four saturation measurements~\cite{Vetsch2010}.

The measurement near resonance is compromised for a number of reasons and the corresponding signal is not displayed in fig.~\ref{fig:phasemeasurement}. 
First, absorption becomes significant near resonance, which not only alters the polarization state as described by the prefactor of $S_3/S_0$ in eq.~(\ref{Eq_stokes}), but also results in a loss of signal simply because less light is available for detection, particularly in the high optical density scenario examined here. 
Second, the absorptive interaction results in scattering events, which perturbs the initial state preparation of the ensemble due to both optical pumping, and the loss of atoms from the trap due to heating. 
Third, the spectral shape of the signal is complicated near resonance due to the presence of the trapping fields, which results in inhomogeneous broadening of the resonance via ac-Stark shifting of the Zeeman substates~\cite{Kien2005}. 
This inhomogeneous broadening takes a different form for the two polarization states, further complicating the expected spectrum. 
One may avoid all of these issues, however, by measuring the phase for short time intervals and at significant detunings, such that absorption and optical pumping is negligible. 

The functional form used for the fit in fig.~\ref{fig:phasemeasurement} is derived from the complex transfer function for the transmitted light given by $\exp(i  (\tilde{n}-1) 2 \pi l/\lambda)$, where $\tilde{n}$ is the complex refractive index (defined, e.g., in~\cite{Sobelman1979}), $l$ is the length of the sample and $\lambda$ is the probe wavelength. 
The dispersive component yields the phase shift of the probe light 
\begin{equation}
\phi(\omega)  = \frac{2 \pi l}{\lambda}\mathrm{Re}\{{\tilde{n}-1}\}  =   - \phi_{\mathrm{max}} \sum_{F'} 2 \frac{\sigma_{F'}}{\sigma_{5}} \frac{\Delta_{F'}}{\Delta_{F'}^2+1}, 
\label{Eq_phi}
\end{equation}
where $\Delta_{F'}=2 (\omega-\omega_{F'})/\Gamma$ is the detuning of the probe light angular frequency $\omega$ from the transition resonance angular frequency $\omega_{F'}$, normalized to the HWHM atomic linewidth  $\Gamma/2$. 
We only consider the transitions from the  $F=4$ ground state to each excited state $F'=3,4,5$, which have a natural linewidth of $\Gamma/2 \pi = 5.2$\,MHz.  
We further denote the maximum phase shift due to the $F=4 \rightarrow F'=5$ transition, which occurs at $\Delta_{F'=5} = -1$, by $\phi_{\mathrm{max}}$ and note that  $\phi_{\mathrm{max}}=\OD/4$~\cite{Sobelman1979}, where $\OD$ is the corresponding maximum optical density occurring at $\Delta_{F'=5}=0$. 
Finally, $\sigma_{F'}/\sigma_{5}$ is the theoretical ratio of the effective far-detuned resonant absorption cross-sections ($\sigma_{3}/\sigma_{5}=7/44$ and $\sigma_{4}/\sigma_{5}=21/44$)~\cite{Steck1998}.  

From the fit of eq.~(\ref{Eq_phi}) to the data shown in fig.~\ref{fig:phasemeasurement}, we are able to quantitatively characterize the optical interface. 
The fitted maximum phase shift of $\parmode$ is $\phi_{\mathrm{max},\|} = (6.98\pm0.02)$\,rad. 
At far blue-detunings, $\phi_{\|} (\omega) \propto N_{\rm{at}}/[ (\omega_{F'}-\omega)/ 2 \pi]$ 
with a proportionality constant of $\phi_{\mathrm{max},\|}\times (\Gamma/2 \pi) /1021 = (36\pm2)$\,mrad\,MHz/atom. 
The corresponding constant for the phase difference $\Delta \phi$ is $(23\pm1)$\,mrad\,MHz/atom. 
The maximum optical density inferred from  $\phi_{\mathrm{max},\|}$ is  $\textrm{OD}_{\|} = (27.93\pm0.07)$. 
This corresponds to an optical density per atom of $\eta=\textrm{OD}_{\|} /N_{\rm{at}} = (0.027\pm0.002)$. 
The measured phase shift and optical density per atom are comparable to what has been previously observed with a single trapped rubidium atom coupled to a strongly focussed Gaussian laser beam~\cite{Aljunid2009}. 
However, by comparison our system offers two additional benefits. 
First, it is possible to obtain efficient coupling for many atoms. 
Second, our interferometric measurement relies on polarization modes propagating through the same fiber and is thus immune to common-mode path-length fluctuations.

Our measurement also yields an effective far-detuned resonant absorption cross-section of $\sigma_{5}= \eta A_{\mathrm{eff}}= (0.94\pm0.06)\times 10^{-9}$\,cm$^2$, where $A_{\mathrm{eff}}$ is the effective mode area~\cite{Warken2007}. 
This is about 32\,\% 
less than the accepted value of $1.4\times 10^{-9}$\,cm$^2$ for the $F=4 \rightarrow F'=5$ transition of the D2 line of cesium~\cite{Steck1998}. 
This deviation could be due to unintended optical pumping in the trap-loading process or imperfect control of the polarization of the probe or trapping beams.  
The zero-crossing of the fit is $(-4.6\pm0.1)$\,MHz, which is approximately 10\,MHz lower than the value expected from our previous work~\cite{Vetsch2010}. 
We attribute this to an unintended offset in the frequency lock of the laser, which might also lead to a systematic error of up to 10\,$\%$ in the determination of the value of the maximum phase, $\phi_{\mathrm{max},\|}$. 

\begin{figure}
\includegraphics[width=0.45\textwidth]{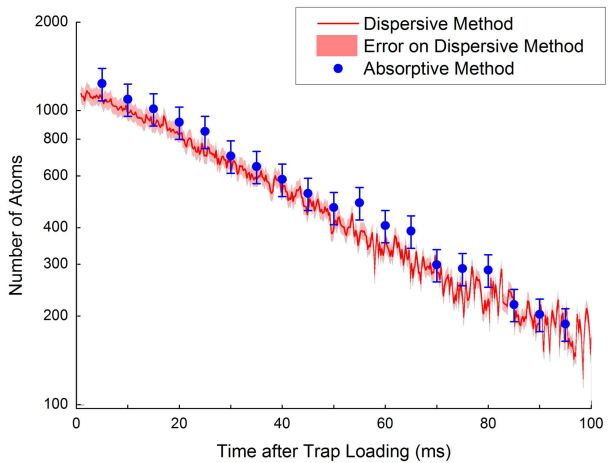}
\caption{ \label{fig:QNDmeasurement}
Red line: Continuous atom number measurement derived from the phase shift (128 averages) scaled using the quantity $(23\pm1)$\,mrad\,MHz/atom. 
Blue dots: Pulsed absorption measurements (16 averages) scaled by a preceding saturation-based atom number measurement (typical uncertainty $\sim$12\%).
}
\end{figure}

The dispersive nature of this detection scheme permits atom number measurements far off-resonance, where absorption is negligible and scattering rates are low. 
In order to demonstrate that this technique can be used without introducing significant additional heating into the system, we performed a continuous atom number measurement of the atomic ensemble derived from the phase and compared it with a series of pulsed resonant absorption measurements with successively longer delays (see fig.~\ref{fig:QNDmeasurement}). 
The pulsed absorption measurements reveal an exponential decay of the number of trapped atoms with a time constant of $(48\pm4)$\,ms. 
The continuous phase signal also decays in $(48\pm1)$\,ms, which indicates that no significant additional loss is introduced by the presence of the dispersive probe beam. 
In this measurement, the probe beam had a power of about 5\,pW and was blue-detuned by +165\,MHz from the $F=4 \rightarrow F'=5$ transition, which corresponds to a scattering rate of $\sim$\,50\,Hz per atom (or $\sim$\,5 photons scattered by each atom over the 100\,ms duration of the measurement). 
For detunings of $\lesssim$\,70\,MHz and using the same probe power, the time constant inferred from the phase measurement drops below that of the resonant absorption measurement, despite there still being very low absorption. 
We believe this is due to the probe beam redistributing the initially even populations in the  Zeeman substates by optical pumping.  
Such a process would alter the relative coupling, and therefore the relative phase shift of $\parmode$ and $\perpmode$,  thereby changing the birefringent effect that underlies our technique. 
Indeed, we have observed that under certain conditions that lead to optical pumping, the birefringent signal is reduced to almost zero in a few milliseconds. 

We note that the sensitivity of all measurements presented above was limited by technical detector noise. 
It is therefore not yet clear that our method can reach the shot-noise limit, in particular in the presence of possible time-dependent polarization mode dispersion, e.g., due to Brillouin scattering. 
It is nevertheless instructive to consider the ultimate sensitivity of the method in the absence of such technical noise sources. 
From the measured phase difference of  23\,mrad\,MHz/atom and the minimal detuning of 70\,MHz at which the heating of the atoms can be neglected for a probe power of 5 pW, we estimate the shot-noise limited sensitivity of the dispersive atom number measurement to be about 0.7\,Hz$^{-1/2}$. 
For an integration time of 5\,ms, i.e., significantly shorter than the storage time of the trap, we should thus in principle be able to dispersively detect the presence of down to 10--20 atoms. 
Furthermore, by increasing the coupling strength of the atoms to the probe field, e.g., through appropriate optical pumping of the atoms and by reducing their distance to the nanofiber surface, it should even become possible to realize non-destructive measurements at the single atom level. 

In conclusion, we have demonstrated a dispersive optical interface with very efficient coupling corresponding to an effective resonant optical density per atom of $0.027$. 
Moreover, the strong birefringent effect inherent to the system enables a simple and effective readout of the phase which makes it an attractive optical-fiber-based platform for non-destructive measurement and coherent manipulation of the quantum states of cold atoms~\cite{Chaudhury2007,Windpassinger2008}. 

This work was supported by the Volkswagen Foundation and the European Science Foundation. 
We thank P. Schneeweiss for his careful proof reading of the manuscript and helpful comments.

\end{document}